\documentclass[a4paper,11pt]{article}
\usepackage{pos}
\usepackage{xcolor}
\usepackage{subcaption} 
\usepackage{graphicx}   
\usepackage{float}
\usepackage{bbm}
\usepackage{xspace} 
\usepackage{orcidlink}

\newcommand{\br}[1]{\left\langle #1 \right |}
\newcommand{\kt}[1]{\left| #1 \right \rangle}
\newcommand{\Ntot}{N_{\mathrm{tot}}}
\newcommand{\LamSRG}{\Lambda_{\mathrm{SRG}}}
\newcommand{\LamNN}{\Lambda_{\mathrm{NN}}}

\newcommand{\etal}{\textit{et al.}\xspace}

\newcommand{\LiS}{{}^{6} \mathrm{Li}\xspace}
\newcommand{\HeF}{{}^{4} \mathrm{He}\xspace}
\newcommand{\HeT}{{}^{3} \mathrm{He}\xspace}
\newcommand{\HThree}{{}^{3} \mathrm{H}\xspace}
\newcommand{\HT}{{}^{2} \mathrm{H}\xspace}
\newcommand\bv[1]{\vec{#1}}

\newcommand{\MeV}{\mathrm{MeV}}

\newcommand{\fm}{\mathrm{fm}}
\newcommand{\fmin}{\mathrm{fm}^{-1}}

\newcommand{\ChiEFT}{$\chi$EFT\xspace}
\newcommand{\m}[1]{\mathrm{#1}}

\usepackage{ulem}
    
   \renewcommand{\emph}[1]{\textit{#1}}           
   \renewcommand{\emph}[1]{\textit{#1}}           
\title{Scattering Observables from Few-Body Densities and Compton Scattering on $\LiS$}

\author*{Alexander Long\orcidlink{0009-0009-5890-2713}}
\author{Harald W. Grie{\ss}hammer\orcidlink{0000-0002-9953-6512}}

\affiliation{Institute for Nuclear Studies, Department of
Physics,\\George Washington University, Washington DC 20052, USA}

\emailAdd{alexlong@gwu.edu}
\emailAdd{hgrie@gwu.edu}
\abstract{
  The dynamics of scattering on light nuclei is
  numerically expensive using standard methods.
  Fortunately, recent developments allow one to factor the relevant quantities for a given probe into a convolution of an $n$-body Transition Density Amplitude (TDA) and the interaction kernel for a given probe.
  These TDAs depend only on the target, and not the
  probe; they are calculated once for each set of kinematics and can be used for different interactions.\ 
  The kernels depend only on the probe, and not on the target; they can
  be reused for different targets and different kinematics.
  The calculation of TDAs becomes numerically
  difficult for more than four nucleons, but we discuss a new solution through the use of a
  Similarity Renormalization
  Group transformation, and a subsequent back-transformation.
  This technique allows for extending the TDA method to heavier nuclei such as $\LiS$.
  We present preliminary results for Compton scattering on $\LiS$ and
  compare with available data, anticipating an upcoming, more thorough study~\cite{upcoming}.
  We also discuss ongoing extensions to pion-photoproduction and other reactions on light nuclei.
}
\FullConference{The 11th International Workshop on Chiral Dynamics (CD2024)\\
 26-30 August 2024\\
Ruhr University Bochum, Germany\\}

\begin{document}
\maketitle

\section{Introduction}
\begin{figure}[!b]
  \begin{center}
    \includegraphics[scale=0.63]{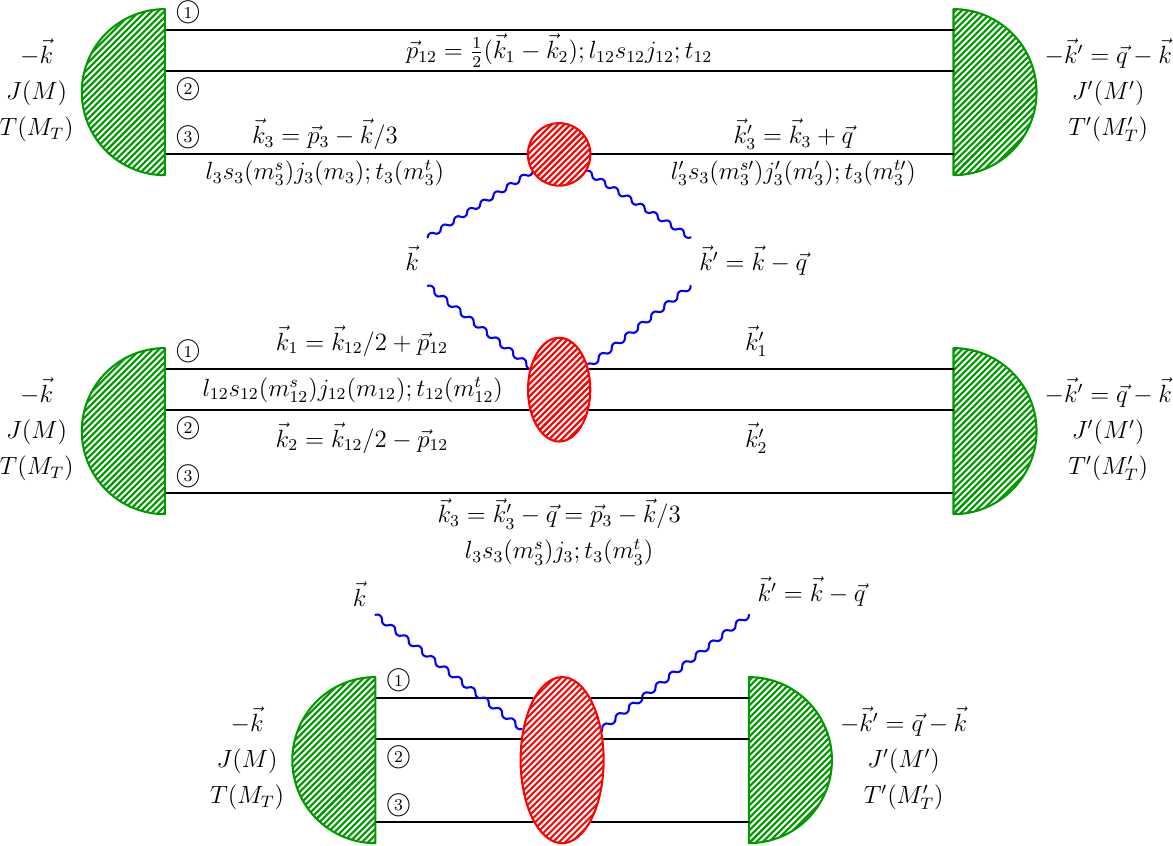}
    \caption{(Color online) Kinematics in the center-of-mass frame and quantum
      numbers for an $A=3$ system in Compton scattering.
      Generalization to other reactions only changes the
      ingoing/outgoing probe; see fig.~\ref{fig:topology}.
      Systems with $A>3$ result in more internal lines
      representing the nucleons.
      Top: one-body processes $\hat{O}_1$ (one active nucleon, two spectators),
      center: two-body
      processes $\hat{O}_{2}$ (two active nucleons, one spectator), bottom: three-body processes
      $\hat{O}_{3}$ (all nucleons active, no spectators). Red - kernel; 
      everything not in red is subsumed into the TDAs;
      green - wavefunction of the nucleons.
      $J(M)$ - spin (projection);  $T(M_T)$ - isospin (projection) of the nucleus. 
      $l_i,s_i(m_i^s),j_i(m_i), t_i(m_i^t)$ - spin, total and orbital angular momentum (and their projections) of the specific subsystem in the kernel being considered.
      From Grie{\ss}hammer \etal \cite{hammer2020}.
  }
    \label{fig:onetwobod}
  \end{center}
\end{figure}

Effective Field Theories (EFTs) in nuclear physics achieve accurate predictions by employing only those
degrees of freedom most pertinent to the physical system
under consideration, rather than relying on the complete set of
degrees of freedom present in the underlying theory.
In this work, we utilize Chiral
Effective Field Theory (\ChiEFT), which adopts nucleons and pions
as its degrees of freedom.
The present study is concerned with the scattering of probes on light nuclei
by means of a Transition Density Amplitude (TDA) ~\cite{hammer2020, Vries2024}.
The TDA method describes the scattering of a probe (such as a photon) on an $A$-nucleon target, 
but we only consider the probe's interaction with $n=1,\ldots,A$ nucleons at a time.
Therefore, we designate the $n$ nucleons with which the probe interacts as \textit{active},
and as \textit{spectators} the background $(A-n)$ nucleons which do not directly interact with the probe.
The active nucleons enter the description of the kernel (along with the probe information), whereas the effect of the spectators on the active nucleons is captured by the TDA.
The $n$-body kernel characterizes the interaction of the probe
with the $n$ active nucleons.
It is irreducible, i.e.~the two-body kernel does not involve the one-body kernel,
and vice versa.
Figure~\ref{fig:onetwobod} provides an illustrative
example for the case $A=3$.

The total scattering amplitude off an $A$-body nucleus is then given by 
\begin{align}
  A_{M}^{M^{\prime} }(\bv{k}, \bv{q})&=
 \langle M^{\prime}|\bigg[
  \binom{A}{1} \hat{O}_{1}(\bv{k}, \bv{q}) +
  \binom{A}{2} \hat{O}_{2}(\bv{k}, \bv{q}) +... + 
  \binom{A}{A} \hat{O}_{A}(\bv{k}, \bv{q})
  \bigg]|M
  \rangle\;\;,
  \label{eq:totalscat}
\end{align}
where $\hat{O}_n$ is the $n$-body kernel, $M,M'$ are the spin projections of the
incoming/outgoing state of the target nucleus, and there are
$\binom{A}{n}$ ways for a probe to interact with $n$ nucleons. 
While this decomposition is \emph{per se} exact and valid for any interaction, computing all TDAs involving up to $A$ active nucleons would be tedious. 
Fortunately, \ChiEFT predicts that $n$-body interactions follow a hierarchy of scales~\cite{WEINBERG1992114}, so that 3-body contributions and higher are negligible  in the first few orders for typical momenta $k\sim m_\pi$ in many processes,
such as Compton scattering, pion photoproduction, and pion scattering.
Therefore, we use only the first two terms of eq.~\eqref{eq:totalscat}. 
In practice, for Compton scattering, this is enough for amplitudes accurate on the 5\% level ~\cite{hammer2020,hammer4He}. 
\section{Kernels and Densities}
The one-body and two-body kernels must be considered separately.
Their form is different, and they require a one- and
a two-body TDA, respectively.
The central result of ref.~\cite{hammer2020} is that one may write the matrix element of $n$-body operators  inside the nucleus much more simply;
in particular, the one-body contribution reduces to
\begin{align}
  \left\langle M^{\prime}\left|\hat{O}_{1}(\bv{q})\right|
  M\right\rangle=\sum_{\substack{m_{3}^{s \prime}\,
  m_{3}^{s}\\m_3^t}}\hat{O}_{1}\left(m_{3}^{s \prime} m_{3}^{s},
  m_{3}^{t} ; \bv{q}\right) \rho_{m_{3}^{s \prime}
  m_{3}^{s}}^{m_3^{t} M_{T}, M^{\prime} M}(\bv{q})\label{onebodyOrig}\;\;.
\end{align}
Here, $\rho$ is the \textit{one-body Transition Density Amplitude} (TDA)
and can be interpreted as the probability amplitude that a nucleon absorbs a
momentum $\bv{q}$ transferred from the projectile to the nucleus and concurrently undergoes a change from the un-primed to the primed quantum numbers; 
see their definition in fig.~\ref{fig:onetwobod}.
The two-body expression equivalent to eq.~\eqref{onebodyOrig} is: 
\begin{align}
  \left\langle M^{\prime}\left|\hat{O}_{2}\right| M\right\rangle &=
  \sum_{\alpha_{12}^{\prime}, \alpha_{12}} \int \mathrm{d} p_{12}\:
  p_{12}^{2} \mathrm{~d} p_{12}^{\prime}\: p_{12}^{\prime 2}\;
  \hat{O}_{2}^{\alpha_{12}^{\prime} \alpha_{12}}\left(p_{12}^{\prime},
  p_{12}\right) \rho_{\alpha_{12}^{\prime} \alpha_{12}}^{M_{T},
  M^{\prime} M}\left(p_{12}^{\prime}, p_{12} ; \bv{q}\right)\label{twobody}\;\;.
\end{align}
Here, a pair of active nucleons absorbs a momentum transfer $\bv{q}$ and changes its quantum numbers from
$\alpha_{12}$ to $ \; \alpha'_{12}$. 
Each represents a set of quantum numbers which fully characterize the nucleon pair, including
$t_{12}$, $m_{12}^t, s_{12}, j_{12},$ and $ l_{12}$ but do not involve spectators;
for more details, see Grie{\ss}hammer~\etal~\cite{hammer2020,hammer4He} and Grie\3hammer's contribution~\cite{Griesshammerproc}.
Just like the one-body case, the two-body TDA can be interpreted as a
transition probability density amplitude.
It depends on the incoming/outgoing quantum numbers
$\alpha_{12}$/$\alpha_{12}'$ of the system of the two active nucleons, and also on their initial and final
relative momenta $p_{12}$ and $p_{12}'$, which are integrated over.
As a result, the file size for the two-body nucleon TDAs is approximately 20 MB
per energy and angle, whereas those of the one-body TDAs are on the order of a few KB.
Importantly, the densities $\rho$ can for a given momentum transfer $\vec{q}$ be computed directly from a nuclear
potential such as the chiral Semilocal Momentum-Space ($\chi$SMS) potential \cite{Reinert2018},
without reference to the interaction kernels $\hat{O}_1$ or $\hat{O}_{2}$.
\section{SRG Transformation}
\subsection{The Method}
Previous work using the TDA formalism analyzed $\HeT$ and $\HeF$~\cite{hammer2020, hammer4He}, but the extension to $\LiS$ is more complicated
than is feasible with currently available computing hardware.
To this end, a \textit{Similarity Renormalization Group} (SRG) transformation~\cite{SRG, Furnstahl2013}
is applied to the potential before the TDA is calculated. Both
with and without an SRG transformation, the TDA calculation requires a parameter $\Lambda$ 
beyond which we assume the potential has no impact on observables.
The choice is not arbitrary; we must choose $\Lambda$ large enough that the 
impact of the potential at momenta above this value on observables is already small, but unfortunately,
nuclear potentials, such as the $\chi$SMS potential \cite{Reinert2018}, do usually
not fall off rapidly at high momenta.
Therefore, a large range of momentum values must be used,
which increases the computational cost.
The SRG transformation is unitary and 
shifts relevant physics into the low-momentum
region, thereby lowering the minimum effective $\Lambda$ needed in the SRG-evolved space.
This significantly improves the convergence rate of calculations
and makes $A\ge6$ actually possible.
An SRG transformation can be thought of as a local averaging or
smoothing of the potential, resulting in decreased ``resolution''
 without losing any of the underlying information at low momenta (i.e.~low resolution) or compromising its physics.
 \begin{figure}[!hb]
  \centering
  \begin{subfigure}{0.45\textwidth}
    \centering
    \includegraphics[width=\linewidth]{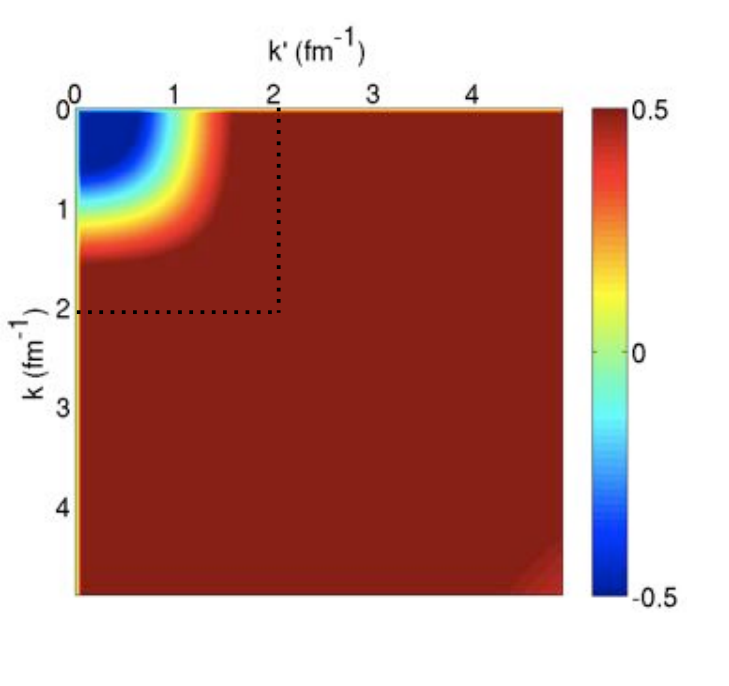}
    \caption{High Resolution (before much SRG is applied). }
    \label{fig:highres}
  \end{subfigure}
  \hfill
  \begin{subfigure}{0.45\textwidth}
    \centering
    \includegraphics[width=\linewidth]{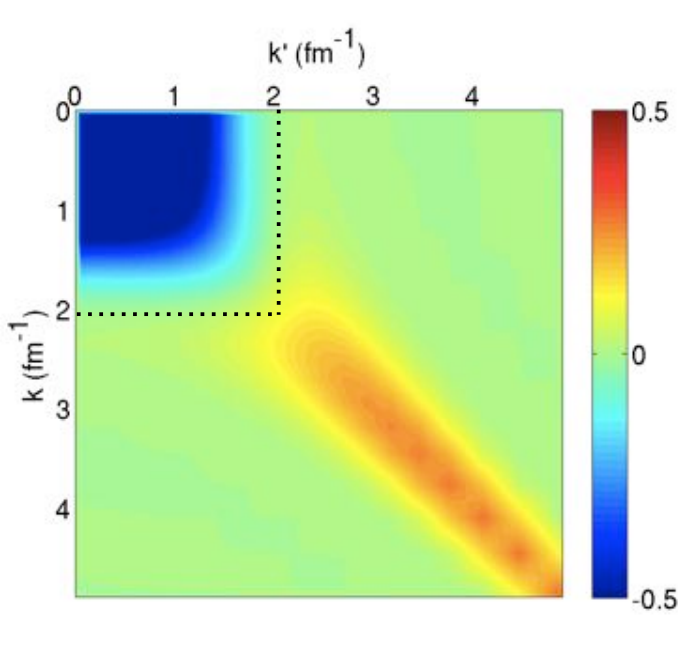}
    \caption{Low Resolution (evolved).}
    \label{fig:lowres}
  \end{subfigure}
  \caption{(Color online)
    Nuclear potentials $V(k,k')$.
    Figure with permission from
    Furnstahl \etal \cite{Furnstahl2013}.~Based on a figure by K.~Hebeler \cite{Hebeler2022}.
  }
  \label{fig:SRGtransform}
\end{figure}
In the under-evolved, high resolution panel, fig.~\ref{fig:highres}, the potential does 
not go to zero rapidly at large momenta, whereas it does once the
transformation is applied in the right panel; see fig.~\ref{fig:lowres}. As a
result, a cutoff can be made at $\Lambda\simeq2 \mathrm{fm}^{-1}$ without
losing much accuracy, whereas the under-evolved potential requires at least $\Lambda\simeq5 \mathrm{fm}^{-1}$.
The calculation time is proportional, at minimum, to the number of array elements present;
therefore we gain at least a factor of $(5/2)^2=6.25$ in efficiency.
In practice, the gains are even higher because near-zero values of the potential at large momenta mean sparse grids can be used.

The SRG transformation is essential, but it also creates a
change in the physical meaning of the free variables.
In fact, any unitary transformation ($U^\dag U =\mathbbm{1}$) also transforms the coordinates.
For simplicity, consider a transformation on a two nucleon potential:
\begin{align}
  V(p\,p)=
  \br{p'}V\kt{p} 
  = \br{p'} U^\dag\left( U V U^\dag\right)\;U
  \kt{p}
  = \br{\widetilde{p}\,'} V_{\mathrm{SRG}}
  \kt{\widetilde{p}}=V_{\mathrm{SRG}}(\widetilde{p},\widetilde{p}\,')\label{SRG1}
\end{align}
So, referring to the free variables in an SRG-transformed potential as
``momenta'' is to some extent incorrect,
as they are not eigenstates of the physical momentum operator.
The Lagrangeans that generate the Feynman diagrams in the kernel, however,
depend on physical momenta, so we cannot directly use an SRG-evolved 
TDA with a non-SRG-evolved kernel.
To solve this, previous work with SRG transformations transformed
the kernels into the SRG-evolved space.
However, in the TDA formalism, this method would require
giving the kernels information about the SRG transformation,
thereby breaking kernel-density independence. 
The SRG transformation can take many different forms~\cite{SRG, Furnstahl2013};
we wish to allow for this and for future developments without having to create a new kernel transformation prescription each time.
Therefore, we developed a method where the SRG transformation is performed on the potential,
then the densities are computed in the SRG-evolved space, 
before finally we apply an inverse SRG transformation to the densities, so that the resulting TDA is in the space of physical momenta and may be used directly with un-evolved kernels~\cite{XiangXiang}.

After the SRG transformation, the TDA calculations for $A\ge6$ use the No-Core Shell Method (NCSM)~\cite{Zhan2004};
this involves the selection of a characteristic width $\omega_H$ and an expansion in basis states.
If that basis is infinite-dimensional, it 
forms a complete set,
but we truncate it at the harmonic oscillator state $\Ntot$,
and estimate convergence by comparing calculations for different values of $\Ntot$.
The parameter $\LamSRG\in [0,\infty[$ represents the progression of the SRG-evolution of the potential, as seen in fig.~\ref{fig:SRGtransform}, with $\LamSRG=\infty$ corresponding to no evolution.
$\LamSRG$, $\Ntot$ and $\omega_H$ can all affect observables.
During the TDA calculation, we obtain the binding energy of the simulated system
and choose $\omega_H$ such that it most closely corresponds to the experimental value, whereas $\Ntot$ is taken as large as feasible.

\begin{figure}[!t]
  \begin{center}
    \includegraphics[width=\linewidth]{
    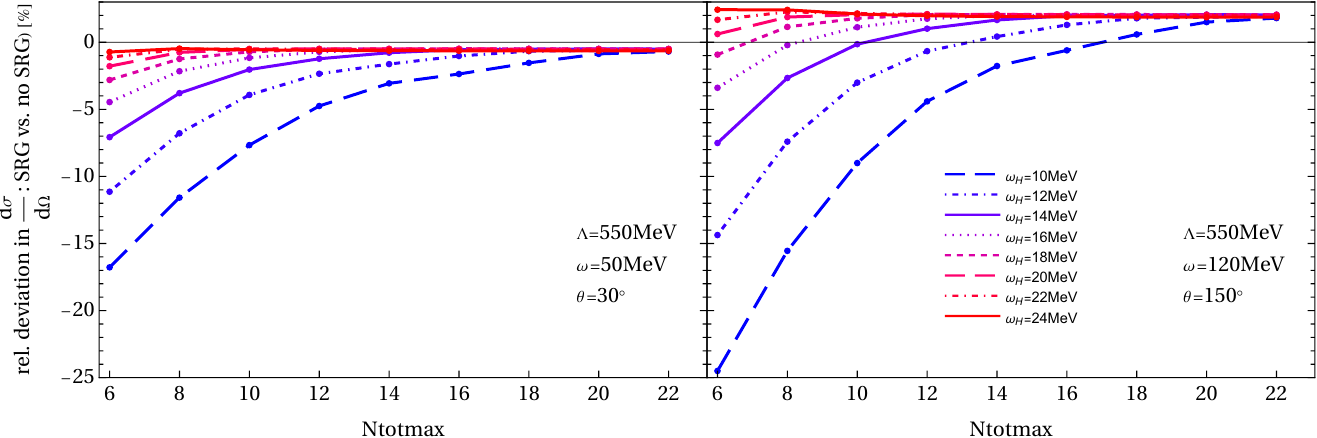}
    \caption{(Color online)  SRG convergence of the $\HeF$\, Compton scattering cross section: exact vs.~SRG-evolved for $\LamSRG=1.88 \fmin$. Left: low energy and low momentum transfer (forward angles); right: higher energy and momentum transfer (back angles). 
    Deviations are due to neglected many-body interactions and truncation in $\Ntot$.
  ``Relative deviation of SRG result from result without SRG''$:= \frac{\mathrm{SRG}}{\mathrm{no~SRG}}-1$, in per-cent.
}
    \label{fig:SRGConverge4He}
  \end{center}
\end{figure}

Equation \eqref{SRG1} is, strictly speaking, incorrect since the SRG transformation induces many-body interactions.
Our methodology accounts for the 2- and 3-body interactions (including the induced ones), but 4-body interactions and higher are neglected.
As a result, our SRG transformation and back transformation are not strictly unitary, and it is essential to test the effect of this approximation.
$\HeF$ is the largest system where a TDA can be calculated with ease without an SRG transformation,
so we compare Compton scattering on $\HeF$ both with and without the SRG transformation
before moving to the more involved $\LiS$, where we only have access to the SRG-transformed TDAs.
Fortunately, fig.~\ref{fig:SRGConverge4He} shows that for $\HeF$ the uncertainty associated with these induced interactions
is small ($\leq2\%$) for even the furthest SRG transformation we considered ($\LamSRG=1.88 \fmin$). 
This holds for small momentum transfers, where we observe that on-diagonal matrix elements dominate, and for large momentum transfers, where off-diagonal matrix elements are very important. This also  tests the $\LiS$ NCSM wave function much more than a computation of the binding energy alone.
The small deviation at high $\Ntot$ is due to the induced many-body interactions along
with the fact that the non-transformed result uses the Faddeev method, while the SRG-transformed result uses the NCSM. 
We will carefully assess the extent to which different $\LamSRG$ values affect results~\cite{upcoming}.

\subsection[Results for Compton Scattering on Lithium 6]{Results for Compton Scattering on $\LiS$}
$\LiS$ is a stable solid at room temperature, and 
is therefore relatively simple to conduct an experiment on~\cite{60MeV,86MeV},
but to date there has been no theoretical description of Compton scattering on $\LiS$.
Here, we calculate Compton scattering on $\LiS$ up to and including $\mathcal{O}(e^2 \delta^3) [\m{N^3LO}]$ with expansion parameter $\delta\approx0.4$ \cite{hammer4He} in ``$\Delta(1232)$-ful'' \ChiEFT. This involves the same kernels used
in Grie{\ss}hammer \etal~\cite{hammer2020, hammer4He} and the same central values of the isoscalar (proton-neutron averaged) electromagnetic polarisabilities from the present ``best'' determination, 
$\alpha_{E1}=11.1\times 10^{-4} \fm^3,\;\beta_{M1}=3.4\times10^{-4}\fm^3$; see e.g.~\cite{Myers:2014ace}.  Grie\3hammer's~\cite{Griesshammerproc} contribution provides more detail.
Future work will provide a full uncertainty analysis, but for this 
preliminary presentation, an overall $\pm10\%$ uncertainty (i.e.~$1\sigma=10\%$) is assumed. 
This is consistent with Grie{\ss}hammer \etal~\cite{hammer2020, hammer4He}, as we expect the uncertainty for $\LiS$ to be similar to that of $\HeT$ and $\HeF$, which in turn is primarily due to the
truncation of the chiral Lagrangean.
\begin{figure}[H]
  \centering
  \includegraphics[scale=0.5]{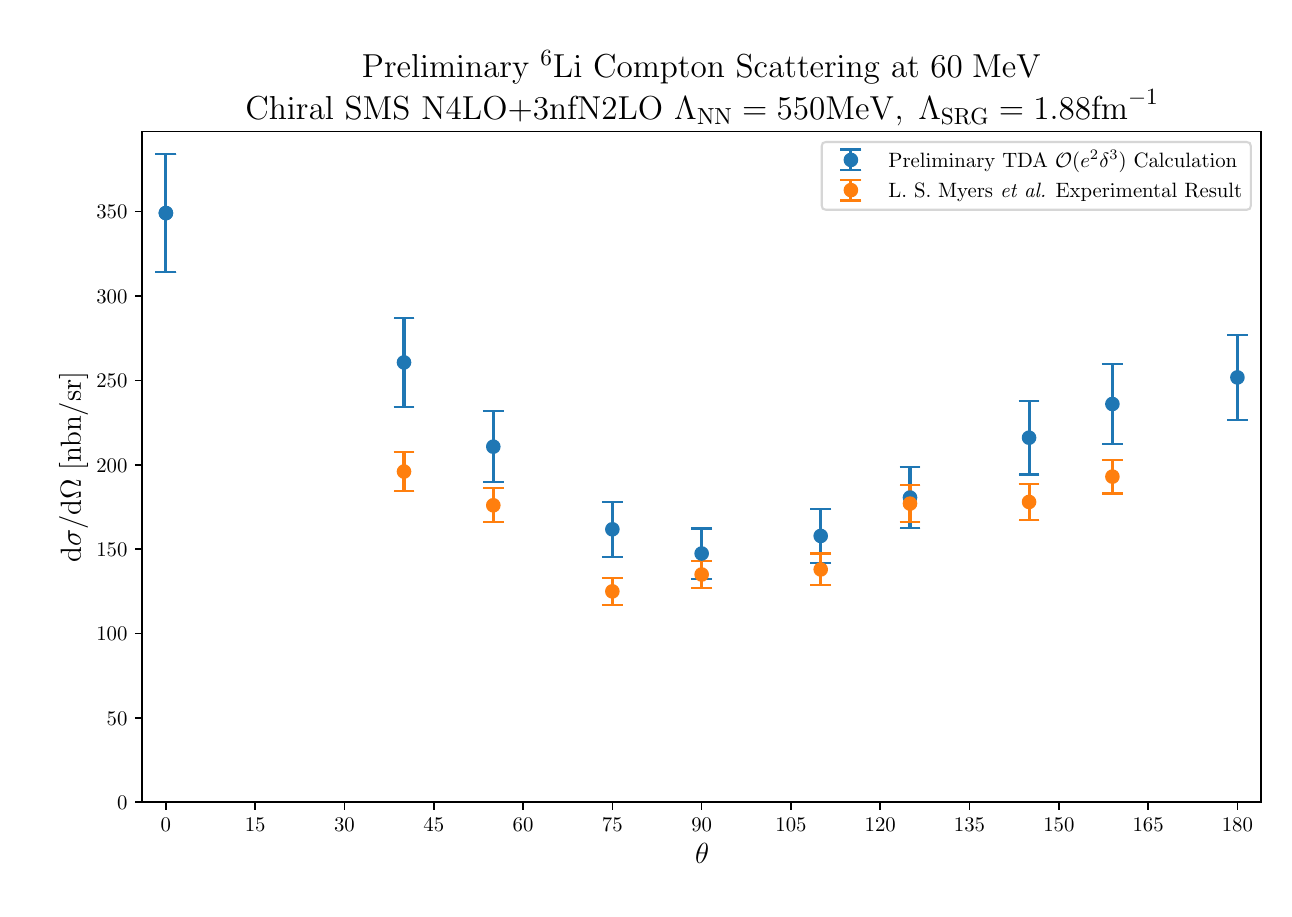}
  \caption{(Color online) Preliminary results for the angle-dependent cross section of Compton scattering on $\LiS$ at $\omega=60\;\MeV$, compared to Myers \etal~\cite{60MeV}.
  }
  \label{fig:TheoryVSExperiment}
\end{figure}
Our preliminary findings in fig.~\ref{fig:TheoryVSExperiment} are promising but leave room for improvement.
The overall angle-dependence of the TDA calculation matches that of the HI$\gamma$S experiment~\cite{60MeV}, but our values appear to differ by a constant offset. 
In an upcoming article~\cite{upcoming}, we will conduct a full uncertainty analysis, including investigation of different $\LamNN$ and $\LamSRG$ values along with 
order-by-order theory convergence.
There are also data at $86\, \MeV$~\cite{86MeV}, which may reveal
the degree to which the apparent normalization discrepancy is energy dependent.
Lastly, we must note that experimental normalization issues cannot be ruled out.

\section{Using TDAs in Different Processes}
\begin{figure}[H]
\centering
\includegraphics[scale=0.5]{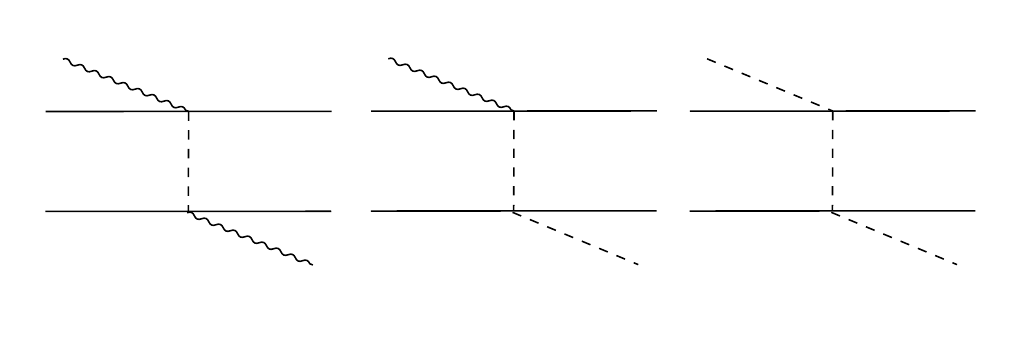}
\caption{Topologically identical contributions to the two-body kernels of Compton scattering, pion-photoproduction and pion scattering.}
\label{fig:topology}
\end{figure}

Having computed TDAs for Compton scattering fr a variety of nuclei and kinematics, we now seek to reuse them for other processes.
Pion-photoproduction and pion scattering are attractive options because their kernels are remarkably similar to those in Compton scattering.
In fact, if one disregards the type of the incoming/outgoing particles and differences in kinematics, the processes are topologically identical;
see example two-body processes in fig.~\ref{fig:topology}.

\subsection{Pion Photoproduction}
Recall that in the TDA formalism, the kernel breaks down into a one-body and a two-body part.
For the pion-photoproduction one-body kernel, we use the 
single-nucleon process, $\gamma N \to \pi N$, which has
been studied extensively both in \ChiEFT and phenomenologically~\cite{pionphoto,
Rijneveen2021,Workman2012,Briscoe2023}.
Its amplitude can be decomposed into
electric and magnetic multipoles $E_{l\pm}, M_{l\pm}$~\cite{pionphoto},
which have been measured with good accuracy up to high multipolarity~\cite{multipolePionPion}.
The resulting scattering matrices $\mathcal{M}$ are exactly what
enter as $\hat{O}_1$ in eq.~\eqref{onebodyOrig}.
This approach solves a significant problem since the calculation of
the one-body pion-photoproduction kernel
to high accuracy directly in \ChiEFT requires including many terms
in the chiral expansion due to the proximity of the $\Delta(1232)$ resonance.
We will also compare a theoretical prediction of these multipoles by Rijneeven \etal~\cite{Rijneveen2021}
to data in future work \cite{upcomingPionPhoto}.

The two-body kernel does not easily decompose into multipoles; therefore, we describe it via pion-exchange currents in \ChiEFT.
Weinberg and Beane \etal ~\cite{Beanephotoprod,Weinberg1992} derived this kernel at threshold up to next-to-next-to-leading order in the chiral expansion.
Additionally, both the one-body and two-body matrix elements have been analyzed at threshold for $\HeT$ and $\HThree$ by
Lenkewitz \etal~\cite{L2011, L2013} and Braun~\cite{Braun}.
The results are typically given in terms of the form factors
$F_{T/L}^{S\pm V}, F_{T/L}^{(a)}-F_{T/L}^{(b)}$; a full description of these is presented in Lenkewitz \etal~\cite{L2011}.
Note that our results in table~\ref{tab:pionphoto} show only the central values, but upcoming work will provide a thorough uncertainty analysis~\cite{upcomingPionPhoto}.
We use a family of $\chi$SMS potentials, whereas Lenkewitz used the AV18 potential, and Braun uses the Idaho (Entem-Machleidt) potential~\cite{Entem2003}, so we expect that our values differ slightly.
\begin{table}
\begin{tabular}{|llll|}
\hline
\multicolumn{4}{|c|}{${}^3\mathrm{He}$ Pion Photoproduction Form Factors}                                                                                                                                                                                             \\ \hline
\multicolumn{1}{|l|}{}                      & \multicolumn{1}{l|}{Lenkewitz \etal AV18~\cite{L2011}} & \multicolumn{1}{l|}{Braun~\cite{Braun}} & $\chi$SMS TDA~\cite{upcomingPionPhoto} \\ \hline\hline
\multicolumn{1}{|l|}{One-body}              & \multicolumn{1}{l|}{}                                                                   & \multicolumn{1}{l|}{}                                    &                                                                    \\ \hline
\multicolumn{1}{|l|}{$F_T^{S+V}$}           & \multicolumn{1}{l|}{0.017(13)(3)}                                                       & \multicolumn{1}{l|}{0.041(2)}                            & -0.017$\pm${uncertainty t.b.a.}                                   \\
\multicolumn{1}{|l|}{$F_T^{S-V}$}           & \multicolumn{1}{l|}{1.480(26)(3)}                                                       & \multicolumn{1}{l|}{1.544(77)}                           & 1.48 $\pm${uncertainty t.b.a.}                                    \\
\multicolumn{1}{|l|}{$F_L^{S+V}$}           & \multicolumn{1}{l|}{-0.079(14)(8)}                                                      & \multicolumn{1}{l|}{}                                    & -0.005 $\pm${uncertainty t.b.a.}                                  \\
\multicolumn{1}{|l|}{$F_L^{S-V}$}           & \multicolumn{1}{l|}{1.479(26)(8)}                                                       & \multicolumn{1}{l|}{}                                    & 1.48 $\pm$uncertainty t.b.a.                                    \\ \hline\hline
\multicolumn{1}{|l|}{Two-body}              & \multicolumn{1}{l|}{}                                                                   & \multicolumn{1}{l|}{}                                    &                                                                    \\ \hline
\multicolumn{1}{|l|}{$F_T^{(a)}-F_T^{(b)}$} & \multicolumn{1}{l|}{-29.3 $\fmin$}                                                      & \multicolumn{1}{l|}{-27.1(33)\;$\fmin$}                           & -29.4 $\fmin$ $\pm${uncertainty t.b.a.}                           \\
\multicolumn{1}{|l|}{$F_L^{(a)}-F_L^{(b)}$} & \multicolumn{1}{l|}{-22.9 $\fmin$}                                                      & \multicolumn{1}{l|}{}                                    & -22.9 $\fmin$ $\pm${uncertainty t.b.a.}                           \\ \hline
\end{tabular}
\caption{$\HeT$ pion photoproduction form factor central values. Full analysis of uncertainty in Long \etal~\cite{upcomingPionPhoto}. \label{tab:pionphoto}}
\end{table}
The values $F_{T/L}^{S+V}$ are small and come from the addition of two much larger numbers with opposite sign;
therefore, the fact that the \textit{relative} mismatch is large is not of concern since the absolute difference between them is small.
Braun \etal~\cite{Braun} analyzed pion photoproduction on $\HT,\HThree,\HeT$ and $\LiS$ with the No-Core Shell Model.
Our work will complement and extend this with a thorough analysis of theory uncertainties for a range of chiral potentials, as well as first steps  above threshold. 

\subsection{Pion Scattering and Future Plans}
In addition to pion-photoproduction, Weinberg and Beane~\etal~\cite{Beane2003, Weinberg1992} developed both the one- and two-body kernels for pion scattering at threshold up to next-to-next-to-leading order in the chiral expansion.
We anticipate extending this analysis above threshold for $\HThree, \HeT, \HeF,$ and $ \LiS$.
Developing both pion-photoproduction and pion-scattering kernels is well underway.
Once complete, we will calculate
these reactions on these targets and compare to data where available.

\section{Conclusion}
We described the Transition Density Amplitude method, a comprehensive framework for
computing scattering observables in light nuclei by factorizing the
amplitude into target‐dependent few-body Transition Density
Amplitudes (TDAs) and probe‐dependent interaction kernels. 
This separation allows us to treat the nuclear structure and the
reaction mechanism independently, which both streamlines the calculation and makes it more efficient.
The central thrust of this work is the successful extension of
the TDA formalism to heavier targets like $\LiS$ by incorporating a
Similarity Renormalization Group (SRG) transformation.
This accelerates the convergence of calculations by
lowering the effective momentum cutoff, and, when combined with an
appropriate inverse transformation, also preserves
the kernel–density independence that is crucial for the versatility
of our approach.
Preliminary results for Compton scattering on
$\LiS$ are not inconsistent with data, but show the need for further analysis which will be addressed in an upcoming publication~\cite{upcoming}. 
We outlined ongoing efforts at extending the formalism to other
reactions, such as pion-photoproduction and pion scattering on light nuclei~\cite{upcomingPionPhoto},
through the development of new kernels both at and above the pion threshold. 
This paves the way for a unified treatment of scattering systems
and ultimately provides a promising route toward
high-accuracy theoretical predictions, deepening our understanding
of nuclear dynamics in light nuclei. 

\subsection*{Acknowledgements}
We thank Andreas Nogga and Xiang-Xiang Sun of FZ J{\"u}lich for collaborating on TDAs and the SRG method, and especially for producing the TDAs on Jureca at the J{\"u}lich Supercomputing Centre (J{\"u}lich, Germany). We also thank G.~Feldman for providing information on $\LiS$ Compton scattering experiments. Additionally, we are grateful to the organizers and participants of Chiral Dynamics 2024 in Bochum.
This work was supported in part by the US Department of Energy under contract DE-SC0015393.
Additional funds were provided by an Enhanced Student Travel Award of the Columbian College of Arts and Sciences of George Washington University. 


\end{document}